\documentclass[twocolumn,english,showpacs,preprintnumbers]{revtex4}

\input{epsf}
\usepackage{epsf}

\def\GRG{{\it Gen. Relativity and Gravitation} }

\def\MNRAS{{\it Mon. Not. R. Ast. Soc.} }

\def\NP{{\it Nucl. Phys.} }
\def\PL{{\it Phys. Lett.} }
\def\PR{{\it Phys. Rev.} }

\def\frac#1#2{{\textstyle{{#1}\over {#2}}}}

\def\lsim{\mathrel{\rlap{\lower4pt\hbox{\hskip1pt$\sim$}}
    \raise1pt\hbox{$<$}}}
\def\gsim{\mathrel{\rlap{\lower4pt\hbox{\hskip1pt$\sim$}}
    \raise1pt\hbox{$>$}}}
\def\sqr#1#2{{\vcenter{\vbox{\hrule height.#2pt
         \hbox{\vrule width.#2pt height#1pt \kern#1pt
         \vrule width.#2pt}
         \hrule height.#2pt}}}}

\def\beq{\begin{equation}}
\def\eeq{\end{equation}}
\def\beqa{\begin{eqnarray}} 
\def\eeqa{\end{eqnarray}}

\def\laq{\raise 0.4 ex \hbox{$<$}\kern -0.8 em\lower 0.62 ex\hbox{$\sim$}}
\def\gaq{\raise 0.4 ex \hbox{$>$}\kern -0.7 em\lower 0.62 ex\hbox{$\sim$}}
\makeatother

\preprint{DF/IST-6.2004}

\begin{document}

\title{The Revival of the Unified Dark Energy-Dark Matter Model ?}

\author{M. C. Bento} \email{bento@sirius.ist.utl.pt}
\affiliation{Departamento de F\'{\i}sica, Instituto Superior T\'{e}cnico,
Av. Rovisco Pais, 1049-001 Lisboa, Portugal}
\author{O. Bertolami} \email{orfeu@cosmos.ist.utl.pt}
\affiliation{Departamento de
F\'{\i}sica, Instituto Superior T\'{e}cnico, Av. Rovisco Pais,
1049-001 Lisboa, Portugal}
\author{A. A. Sen}\email{anjan@cfif3.ist.utl.pt}
\affiliation{Departamento de
F\'{\i}sica and Centro de F\'{\i}sica das Interac\c{c}\~{o}es
Fundamentais, Instituto Superior T\'{e}cnico, Av. Rovisco Pais,
1049-001 Lisboa, Portugal}

\date{\today}

\begin{abstract}
We consider the generalized Chaplygin gas (GCG) proposal for
unification of dark energy and dark matter and show that it admits an
unique decomposition into dark energy and dark matter components once
phantom-like dark energy is excluded. Within this framework, 
we study structure formation and show that difficulties associated to 
unphysical oscillations or blow-up in the matter power spectrum 
can be circumvented. Furthermore, we show
that the dominance of dark energy is related to the time when energy  
density fluctuations start deviating from the linear $\delta \sim a$
 behaviour.

\vskip 0.5cm
\end{abstract}

\maketitle


\section{Introduction}


The GCG model \cite{Kamenshchik,Bento1} is an interesting alternative
to earlier proposals aiming to explain the observed accelerated
expansion of the Universe such as an uncanceled
cosmological constant \cite{Bento2} and quintessence \cite{Ratra}, the
latter being 
a variation of the idea that the cosmological term could
evolve \cite{Bronstein}. 

In the GCG approach one considers an exotic
background fluid, described by the following equation of state

\beq
p_{ch} = - {A \over \rho_{ch}^{\alpha}} ~~,
\label{eqstate}
\eeq
where $A$ and $\alpha$ are positive constants. The case $\alpha=1$
corresponds to the Chaplygin gas. In most phenomenological
analyzes the range $0 < \alpha \le 1$ has been considered. Within the
framework of Friedmann-Robertson-Walker cosmology, this equation of state
leads, after inserted into the relativistic energy conservation equation, to
an energy density evolving as \cite{Bento1}

\beq
\rho_{ch}=  \left[A + {B \over a^{3 (1 + \alpha)}}\right]^{1 \over 1 +
 \alpha}~~,
\label{rhoch}
\eeq
 where $a$ is the scale-factor of the Universe and $B$ an 
integration constant which should be positive for a 
well-defined $\rho_{ch}$ at all times. Hence, we see that at early times
the energy density behaves as matter while at late times it behaves
like a cosmological constant. This dual role is at the heart of the
surprising properties of the GCG model. Moreover, this dependence with
the scale-factor indicates that the GCG model can be interpreted as an
entangled mixture of dark matter and dark energy.

The GCG model has been successfully confronted with various
phenomenological tests: high precision Cosmic Microwave Background
Radiation data \cite{Bento3}, supernova data \cite{Supern}, and
gravitational lensing \cite{Silva}. In a recent work \cite{Bertolami},
it has been explicitly shown that regarding the latest supernova data
\cite{Tonry}, the GCG model is degenerate with a dark energy model
involving a phantom-like equation of state (See also Ref \cite{Tritha}
for a detail study with different Supernova data sets). This excludes the
necessity of invoking an unphysical fluid violating the crucial
dominant energy condition for theoretical model building of our
Universe which leads to the big rip singularity in future. The GCG, on the
other hand, can mimic such an equation of state, but without any such
pathology as asymptotically it approaches to a well-behaved de-Sitter
universe.

Despite all these pleasing features, the main concern with such an
unified model is that it produces unphysical oscillations or even an
exponential blow-up in the matter power spectrum at present
\cite{Sandvik}.  This is expected from the behaviour of the sound
velocity through the GCG fluid. Although, at  early times, the GCG
behaves like  dark matter and its sound velocity is vanishing, as one
approaches the present, the GCG starts behaving like a dark energy
with a substantial negative pressure yielding a large sound velocity
which, in turn,  produces oscillations or blow-up in the power
spectrum. In any unified approach this is unavoidable unless one can
successfully identify the dark matter and the dark energy components
of the fluid. Naturally, these components are interacting as both are
entangled within a single fluid. This is the main motivation of our
present investigation.  We show that the GCG is a unique mixture of
interacting dark matter and a cosmological constant-type dark energy,
once one excludes the possibility of phantom-like dark energy. Due to
the interaction between the components, there is a flow of energy from
dark matter to  dark energy. This energy
transfer is vanishingly small until recent past, resulting in a
negligible contribution from the cosmological constant at the time of
gravitational collapse ($z_c \simeq 10$). This makes the model
indistinguishable from a CDM dominated Universe in the past. Just
before the present ($z \simeq 2$), the interaction starts to
kick off producing a large energy transfer from dark matter to dark
energy leading to its dominance at present. We have also shown that
the epoch of this dark energy dominance is similar to that when dark
matter perturbations start deviating from its linear behaviour.
Moreover, in this approach, the Newtonian equations for small scale
perturbations for dark matter do not involve any $k$-dependent
term; hence, neither oscillations nor blow-up in the power spectrum  are
expected. We should mention that the decaying
dark matter model has been previously considered as an interesting
possibility to solve, within the CDM model, the problem of
over-production of dwarf galaxies as well as the over-concentration of
dark matter in halos \cite{Cen}. Our results show that the GCG model
is an interesting option for that scenario, such that the decay
product is nothing but a cosmological constant.


\section{Decomposition of the GCG fluid}


In Ref. \cite{Bento1}, it has been shown that the GCG Lagrangian density has
the form of a \textit{generalized} Born-Infeld theory:

\beq
{\cal L}_{GBI} = - A^{1 \over 1 + \alpha} 
\left[1 - (g^{\mu \nu} \theta_{, \mu} 
\theta_{, \nu})^{1 + \alpha \over 2\alpha}\right]^{\alpha \over 1 + \alpha}~~,
\label{GenBornInfeld}
\eeq
 which clearly reproduces the Born-Infeld Lagrangian density for
$\alpha = 1$. The field $\theta$ corresponds to the phase of a complex
scalar field \cite{Bento1}.

Let us now discuss the decomposition of the GCG into components. Using
Eqs. (\ref{eqstate}), (\ref{rhoch}) and introducing the redshift
dependence, the pressure is given by

\beq
p_{ch} = -{A \over \left[A+B(1+z)^{3(1+\alpha)}\right]^{\frac{\alpha}
{1+\alpha}}}
\label{totalp}
\eeq
while the total energy density can be written as

\beq
\rho_{ch}=\left[A+B(1+z)^{3(1+\alpha)}\right]^{\frac{1}{1+\alpha}}~~, 
\label{totalrho}
\eeq
where one has set the present value of the scale-factor, $a_0$, to $1$.

Decomposing the energy density into a pressureless dark matter
component, $\rho _{dm}$, and a dark energy component,
$\rho _{X}$ with an equation of state $w_X$, it follows that the equation of state parameter of the GCG can be written as

\beq
w={p_{ch} \over \rho_{ch}}={p_X \over \rho_{dm}+\rho_X}={w_X\rho_X
  \over 
\rho_{dm}+\rho_X}~~.
\label{espar}
\eeq

Thus, using (\ref{totalp}), (\ref{totalrho}) and (\ref{espar}), one
obtains for $\rho_X$

\beq
\rho_X=-{ \rho_{dm} \over
  1+w_X\left[1+\frac{B}{A}(1+z)^{3(1+\alpha)}\right]}~~.
\label{rhox}
\eeq

It is easy to see that, requiring that $\rho _{X} \ge 0$ leads to the
constraint $w_{X} \le 0$ for early times ($z\gg 1$) and $w_{X} \le -1$ for future ($z=-1$). Hence, one concludes that $w_{X} \le -1$ for the entire history of the universe.  The case $w_{X}<-1$
corresponds to the so-called phantom-like dark energy, which violates
the dominant-energy condition and leads to an ill defined sound
velocity (see however \cite{Bertolami}). If one excludes this
possibility, then the energy density can be split in an unique way:

\begin{equation}
\rho=\rho_{dm}+\rho_\Lambda  \label{split}
\end{equation}
where

\beq \rho_{dm}={B (1+z)^{3(1 + \alpha)} \over \left[A + B
(1+z)^{3(1+\alpha)}\right]^{\frac{\alpha}{1+\alpha}}}~~,
\label{rhom}
\eeq
and

\beq \rho_\Lambda=-p_\Lambda={A \over
\left[A+B(1+z)^{3(1+\alpha)}\right]^{\frac{\alpha}{1+\alpha}}}~~,
\label{rholam}
\eeq
from which one obtains the scaling behaviour of the energy densities

\beq
{\rho_{dm} \over \rho_\Lambda}={B \over A} (1+z)^{3(1+\alpha)}~~.
\label{scale}
\eeq

\begin{figure}
\epsfxsize 2.5in
\epsfbox{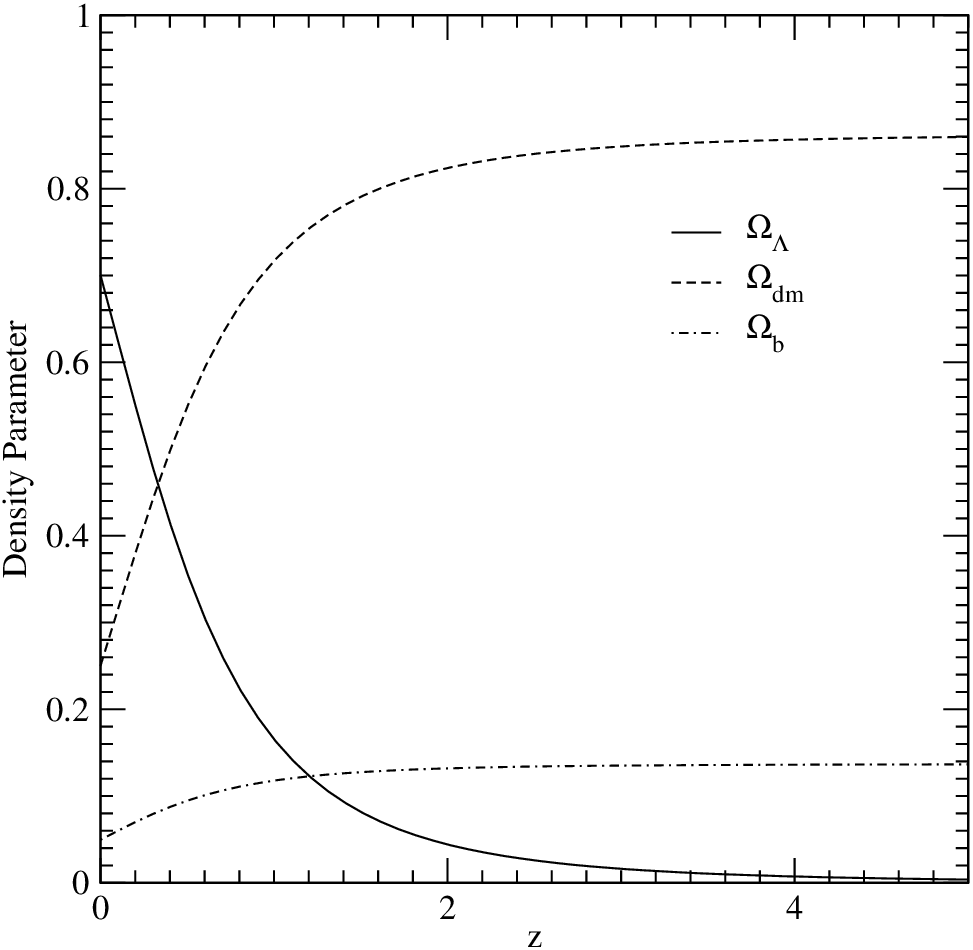}
\caption{\label{fig:fig1} $\Omega_{dm}$ and $\Omega_\Lambda$ and
 $\Omega_{b}$ as a function of redshift. We have assumed
 $\Omega_{dm0}=.25$, $\Omega_{\Lambda0}=0.7$ and $\Omega_{b0}=0.05$ 
 and $\alpha =0.2$.}
\end{figure}

One should note that this type decomposition based on the tachyon-like lagrangian has previously been considered by Padmanabhan and Choudhury \cite{Paddy}.
Next we express parameters A
and B in terms of cosmological observables. From Eqs. (\ref{rhom}) and (\ref
{rholam}), it follows that

\beq
\rho_{ch0}=\rho_{dm0}+\rho_{\Lambda 0}= (A+B)^{\frac{1}{1+\alpha}}~~,
\label{rel}
\eeq 
where $\rho _{ch0}$, $\rho _{dm0}$ and $\rho_{\Lambda 0}$ 
are the present values of $\rho_{ch}$, $\rho_{m}$ and
$\rho _{\Lambda}$, respectively. Parameters $A$ and $B$ can then be
written as a function of $\rho _{ch0}$

\beq
A=\rho_{\Lambda 0}~\rho_{ch0}^\alpha~; \quad
B=\rho_{dm0}~\rho_{ch0}^\alpha~~.
\label{AB}
\eeq

It is also useful to express $A$ and $B$ in terms of $\Omega _{dm0}$,
$\Omega_{\Lambda 0}$, the present values of the fractional energy
densities $\Omega _{dm(\Lambda )}={\rho _{m(\Lambda )}/\rho _{c}}$
where $\rho _{c}$ is the critical energy density, $\rho _{c}= 3H^2/8
\pi G$. Using the Friedmann equation

\beq
3H^2  = 8 \pi G \left[A + B 
(1+z)^{3(1+\alpha)}\right]^{\frac{1}{1+\alpha}}
+ 8\pi G\rho_{b0}(1+z)^3
\label{h2}
\eeq
where $\rho_{b0}$ is the present baryon energy density, one obtains

\beq
A \simeq \Omega_{\Lambda 0}~ \rho_{c 0}^{(1+\alpha)}, ~
B \simeq \Omega_{dm 0}~ \rho_{c 0}^{(1+\alpha)}~~.
\label{params}
\eeq

Hence, with 

\beq
H^2= H_0^2 \left[\left[\Omega_{\Lambda 0} + \Omega_{dm 0} (1+z)^{3(1+\alpha)}
\right]^{\frac{1}{1+\alpha}} + \Omega_{b0}(1+z)^{3}\right]~~
\label{hubble2}
\eeq
one can express the fractional energy
densities $\Omega _{dm},~\Omega _{\Lambda }$  and $\Omega_{b}$ as

\beqa
\Omega_{dm} & = & {\Omega_{dm0}(1+z)^{3(1+\alpha)} \over
\left[\Omega_{\Lambda
    0}+\Omega_{dm0}(1+z)^{3(1+\alpha)}\right]^{\alpha/
(1+\alpha)}X}\\
\Omega_\Lambda & = &{\Omega_{\Lambda 0} \over
\left[\Omega_{\Lambda
    0}+\Omega_{dm0}(1+z)^{3(1+\alpha)}\right]^{\alpha/
(1+\alpha)}X}\\
\Omega_b & = & {\Omega_{b0}(1+z)^{3} \over X}\\
\label{rholam2}
\eeqa
where
\beq
 X=\left[\Omega_{\Lambda
 0}+\Omega_{m0}(1+z)^{3(1+\alpha)}\right]^{1/(1+\alpha)}+\Omega_{b0}(1+z)^3~.
\eeq

Finally, given that $\Omega_{dm0}$ and $\Omega_{\Lambda 0}$ are order
one quantities, one sees that at the time of nucleosynthesis, 
$\Omega_\Lambda$ is negligibly small, making the model consistent with
the nucleosynthesis process.

Notice that there is an explicit interaction between dark matter and
dark energy. This can be seen from the energy conservation equation,
which in terms of the components can be written as

\beq
\dot \rho_{dm} + 3 H \rho_{dm} = - \dot \rho_\Lambda ~~.
\label{evorho}
\eeq
 Hence the evolution of dark energy and dark matter are linked so
that energy is exchanged between these components (see
Refs.\cite{Amendola1,Pavon} for earlier work on the interaction
between dark matter and dark energy). One can see from Figure 1, that
until $z \simeq 2$, there is practically no exchange of energy and the
$\Lambda$ term is approximatelly zero.  However, around $z \simeq 2$, the
interaction starts to kick off, resulting in an important growth of the
$\Lambda$ term at the expense of the dark matter energy
density. Around $z \simeq 0.2$, dark energy starts dominating the
energy content of Universe. Obviously, these redshift values depend on
the $\alpha$ parameter and, in Figure 1, we have assumed
$\alpha = 0.2$. Nevertheless, the main conclusion is that in this
unified model, the interaction between dark matter and  dark
energy is practically zero for almost the entire history of the
Universe making it indistinguishable from the CDM model.  The energy
transfer starts just in the recent past resulting in a significant
energy transfer from dark matter to the $\Lambda$-like dark energy. In
the next section we shall see that this epoch of energy transfer is
similar to the one when dark matter perturbations start departing from
its linear behaviour.


\section{ Structure Formation}

In order to study structure formation, it is convenient to write the
0-0 component of  Einstein's equation as

\beq
3H^{2} = 8\pi G (\rho_{dm} + \rho_{b})+ \Lambda ~~,
\label{G00}
\eeq
where $\Lambda$ is given by

\beq
\Lambda = 8\pi G \rho_{\Lambda}~~.
\eeq

The energy conservation equation for the background fluid is given by
Eq.~(\ref{evorho}).  This is reminiscent of earlier work on varying $\Lambda$
cosmology \cite{Bronstein,Freeze,Waga} where the cosmological term
decays into matter particles. In our case, it is the opposite as
$\alpha$ is always positive and hence the energy transfer is from dark
matter to dark energy. This leads to the late time dominance of the
latter and ultimately to the present accelerated expansion of the
Universe.

\begin{figure}
\epsfxsize 2.7in 
\epsfbox{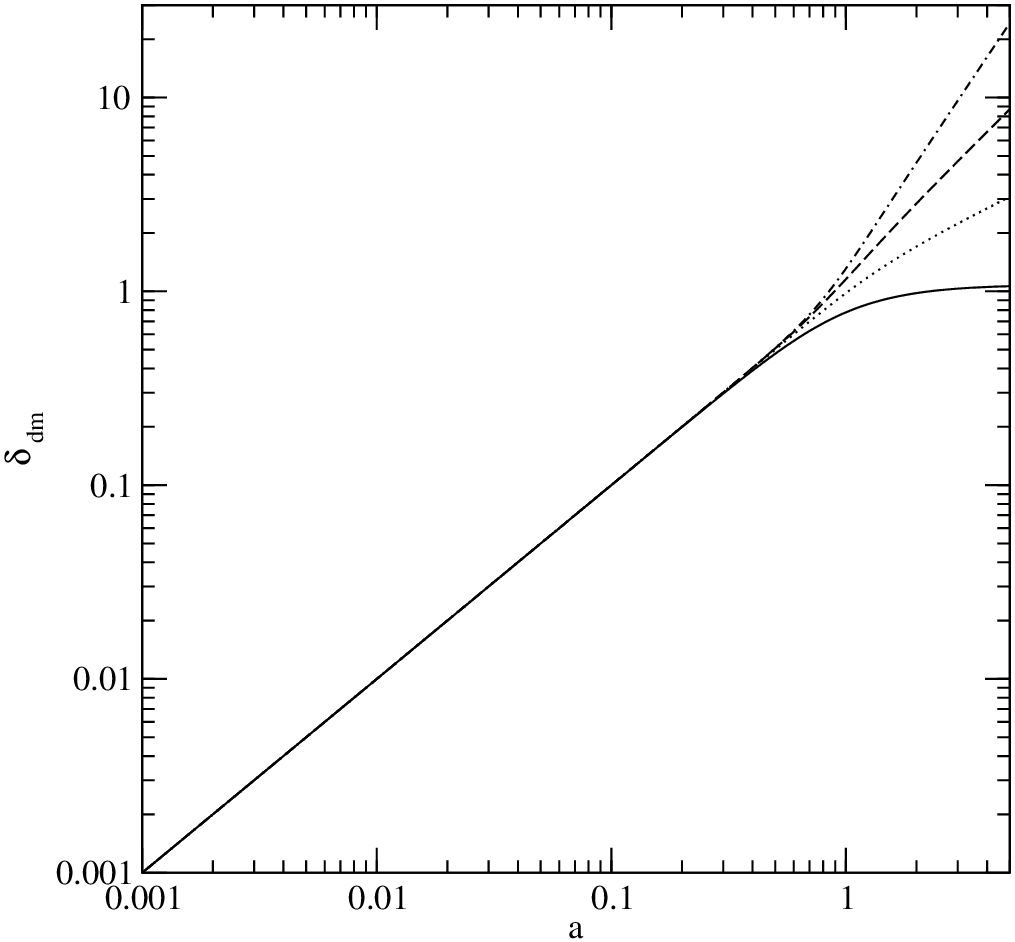}
\caption {$\delta_{dm}$ as function of scale factor.  The solid,
dotted, dashed and dash-dot lines correspond to $\alpha = 0, 0.2, 0.4,
0.6$, respectively. We have assumed $\Omega_{dm}=0.25, \Omega_{b}=0.05$ and
$\Omega_{\Lambda}=0.7$.}
\end{figure}

Let us now consider the issue of energy density perturbations.  We
start by writing down the Newtonian equations for a pressureless fluid
with background density $\rho_{dm}$, and density contrast
$\delta_{dm}$ with a source term due to the energy transfer from dark
matter to the cosmological constant type dark energy. Assuming that
both, the density contrast $\delta_{dm}$ and peculiar velocity $v$ are
small, i.e. that $\delta_{dm} <<1$ and $v << u$, where $u$ is the
velocity of a fluid element of volume, one can write the Euler,
the continuity and the Poisson's equations in the co-moving frame as
follows \cite{Waga}:

\beqa 
\label{eq1}
\ddot{a}x + {\partial{v}\over{\partial{t}}} + {\dot{a}\over{a}}v
= - {\nabla\Phi\over{a}} ~~,\\ 
\label{eq2}
\nabla\cdot v =
-a\left[{\partial{\delta_{dm}}\over{\partial{t}}} +
{\Psi\delta_{dm}\over{\rho_{dm}}}\right] ~~,\\
\label{eq3}
{1\over{a^2}}\nabla^{2}\Phi = 4\pi G \rho_{dm}(1+\delta_{dm}) -
\Lambda ~~, 
\eeqa 
where $\Phi$ is the gravitational potential, and
$\Psi$ is the source term in the continuity equation due to the energy
transfer between dark matter and the cosmological constant type dark
energy. The co-moving coordinate $x$ is related to the proper
coordinate $r$ by $r = ax$. In our case, 

\beq \Psi = -{1\over{8\pi G}}
\dot{\Lambda}~~.  
\eeq

Taking the divergence of Eq.(\ref{eq1}) and using Eqs.(\ref{eq2}) and
(\ref{eq3}), one obtains the small scale linear perturbation
equation for the dark matter in the Newtonian limit:

\beqa {\partial^{2}\delta_{dm}\over{\partial t^2}} +
\left[2{\dot{a}\over{a}} + {\Psi\over{\rho_{dm}}}\right]
{\partial\delta_{dm}\over{\partial t}} &-& \nonumber\\ \left[4\pi
G\rho_{dm} - 2{\dot{a}\over{a}}{\Psi\over{\rho_{dm}}} -
{\partial\over{\partial t}}
\left[{\Psi\over{\rho_{dm}}}\right]\right]\delta_{dm} &=& 0 ~~.  \eeqa

Notice that, if $\Psi = 0$, i.e. without energy transfer,
one recovers the standard equation for the dark matter perturbation in the
$\Lambda$CDM case. One can also check that this occurs for $\alpha
=0$.  Moreover, one can easily see that, in the above equation, there
is no scale dependent term to drive oscillations or blow up in the
power spectrum.

\begin{figure} 
\epsfxsize 2.7in 
\epsfbox{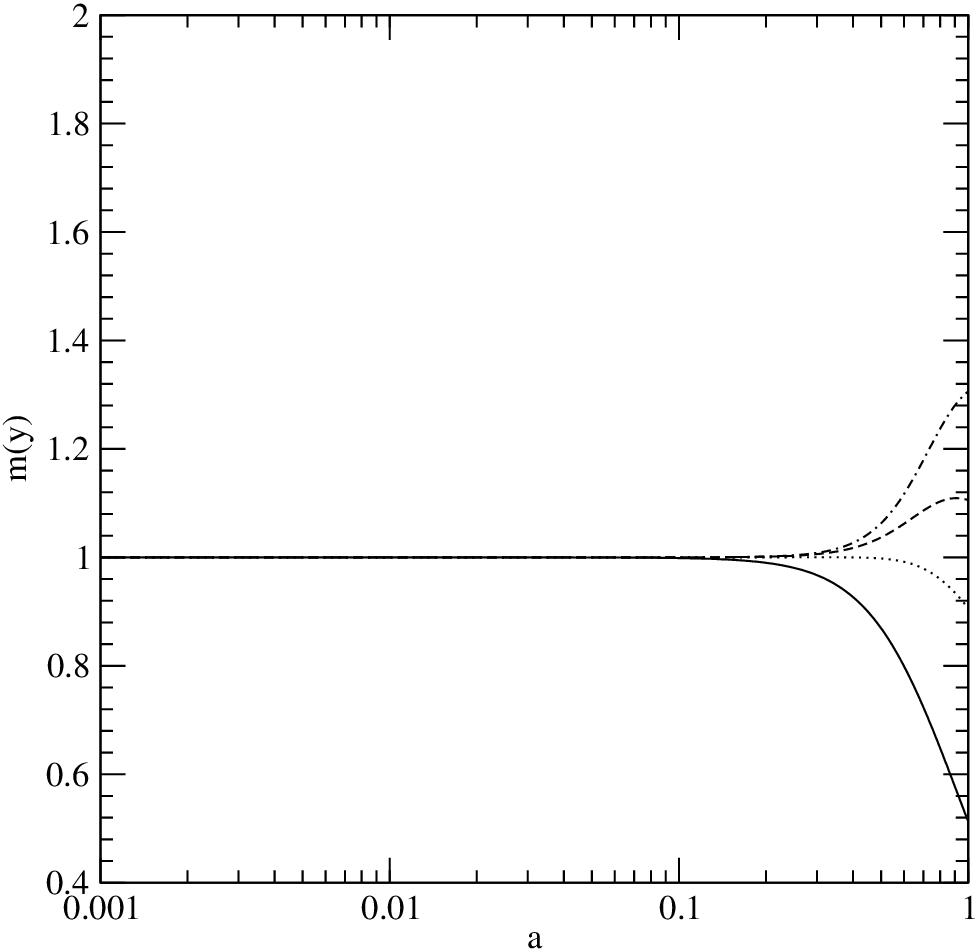}
\caption{The growth factor $m(y)$ as a function of scale factor a.
The solid, dotted, dashed and dash-dot lines correspond to $\alpha =
0, 0.2, 0.4, 0.6$, respectively.  We have assumed $\Omega_{dm}=0.25,
\Omega_{b}=0.05$ and $\Omega_{\Lambda}=0.7$.}
\end{figure}

Let us turn to the evolution for the baryon perturbation in the
Newtonian limit when the scales are well inside the horizon.  Since we
are considering the evolutionary period after decoupling,  the
baryons are no longer coupled to photons, there is no significant
pressure due to Thompson scattering, and one can effectively consider
baryon as a pressureless fluid like the dark matter fluid. Of
course, the interaction between baryons and dark energy is a relevant
issue as it is related to the Equivalence Principle (see
e.g. \cite{Amendola2} and references therein). In what follows we
shall assume that the interaction of dark energy with baryons
vanishes. Thus, as baryons are uncoupled, the strong interaction
between the dark matter and dark matter violates the Equivalence
Principle. Given that it is the parameter $\alpha$ that controls this
interaction ($\alpha =0$ means there is no interaction), it is a
measure of the violation of the Equivalence Principle. One can also
see from the behaviour of $\Psi$, that this violation also starts
rather late in the history of the Universe. In the Newtonian limit,
the evolution of the baryon perturbation after decoupling for scales
well inside the horizon is similar to the one described earlier for dark
matter, but without the source term as there is no energy transfer to
or from baryons. It is given by

\beq {\partial^{2}\delta_{b}\over{\partial t^2}} +
2{\dot{a}\over{a}}{\partial\delta_{b}\over{\partial t}} - 4\pi
G\rho_{dm}\delta_{dm} = 0 ~~, 
\eeq 
where the third term in the l.h.s,
we have dropped the contribution from baryons as it is negligible
compared to the dark matter one.  It is convenient to define for each
component the linear growth function $D(y)$ where $y=\log(a)$,

\beq \delta = D(y)\delta_{0} ~~, 
\eeq 
where $\delta_{0}$ is the
initial density contrast (assuming Gaussian distribution), as well as
the growth exponent $m(y) = D^{'}(y)/D(y)$.

Asymptotically, as dark matter drives the evolution of the baryon
perturbations, hence they grow with the same exponent $m(y)$. However,
their amplitudes may differ and their ratio corresponds to the
so-called  bias parameter, $b \equiv \delta_{b}/ \delta_{dm}$.

 
\begin{figure}
\epsfxsize 2.7in 
\epsfbox{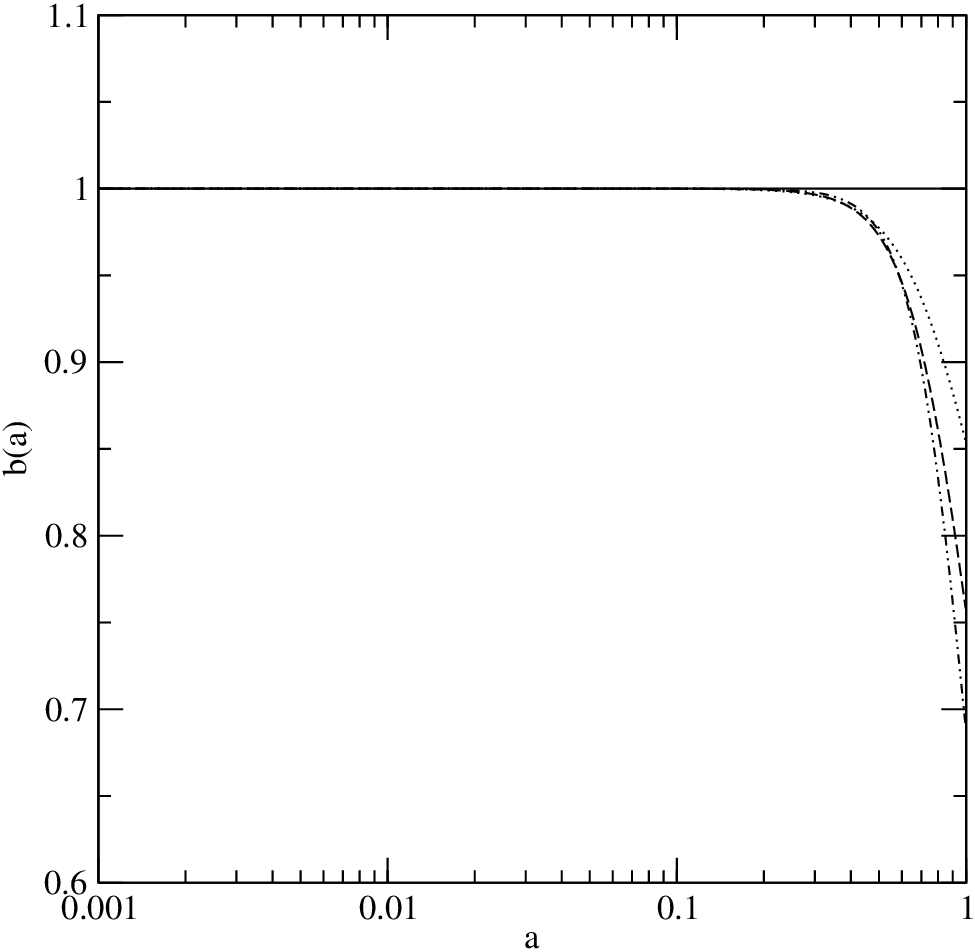}
\caption{The bias $b$ as a function of the scale factor, $a$. The solid, 
dotted, dashed and dash-dot lines correspond to $\alpha = 0, 0.2, 0.4,
0.6$, respectively. 
We have assumed $\Omega_{dm}=0.25, \Omega_{b}=0.05$ and
 $\Omega_{\Lambda}=0.7$.}
\end{figure}

In what follows, we study the behaviour of $\delta_{dm}$, $m(y)$ and
$b$ as function of the scale factor $a$.  While solving the
differential equations for the linear perturbation, the initial
conditions are chosen such that at  $a= 10^{-3}$, the standard linear
solution $D \simeq a$, is reached. In Figure 2, we have plotted the
linear density perturbation for the dark matter, $\delta_{dm}$, as a
function of $\alpha$.  One can see that while for $\alpha = 0$ (the
$\Lambda$CDM case) the perturbation stops growing at late times, for
models with $\alpha > 0$, the
perturbation starts departing from the linear behaviour around a $
\simeq 0.8$ (we have assumed for scale factor at present $a_{0} = 1$)
i.e. $z \simeq 0.25$ which is similar to the epoch when the
$\Lambda$ term starts dominating (cf. Figure 1).  In view of this
behaviour, it is tempting to conjecture that, in our unified model,
the interaction between dark matter and $\Lambda$-like dark energy is
related with structure formation, so that sufficiently high density contrast
($\delta_{dm} >> 1$) can result a large
energy transfer from the dark matter to the dark energy due to which 
acceleration of the Universe sets in.  We should mention that this
kind of scenario has been earlier discussed in
Ref. \cite{Alam}. 
 Thus our study shows that there is a link between
the structure formation scenario and the dominance of dark energy
which ultimately results in the acceleration of the Universe
expansion. This may give possible hint to the solution of so-called
 Cosmic Coincidence problem.

The behaviour of the growth factor $m(y)$ is also quite
interesting. One can see from Figure 3, that between the present and
$z \simeq 5$, the growth factor is quite sensitive to the value of
$\alpha$.  With $\alpha = 0.2$, it increases up to $40\%$ at present
in relation to the value of the $\Lambda$CDM case. Notice  that
$m(y)$ governs the growth of the velocity fluctuations in linear
perturbation theory as the velocity divergence evolves as $\theta = -
H a m \delta_{dm}$; therefore, large deviations of the growth factor
with changing $\alpha$ are detectable via precision measurements of
 large scale structure, through joint measurements of the redshift-space 
power spectrum anisotropy and bi-spectrum from $z = 0$ to $z \simeq 2$.

We have also studied the behaviour of the bias parameter in our model. In
Figure 4, its evolution is shown. The plot suggests that the bias
parameter also changes sharply in recent past with increasing
$\alpha$. This bias extends to all small scales allowing for the
Newtonian limit, being hence distinguishable from the hydrodynamical
or nonlinear bias which takes place only for collapsed objects. Thus,
from the observation of large scale clustering one can distinguish
the non-zero $\alpha$ case from the $\alpha = 0$ ($\Lambda$CDM) case.

One can also see from Figure 2 that there is no suppression of
$\delta_{dm}$ at late times for any positive value of $\alpha$, and
thus one should not expect the corresponding suppression in the power
spectrum normalization, $\sigma_{8}$, for the total matter
distribution. This was one major problem in the previous GCG model
approach which, as pointed out by Sandvik et al. \cite{Sandvik},
cannot be solved even after the inclusion of baryons. In the new approach
we propose here, one can successfully overcome this difficulty.
 
Another interesting cosmological probe for our model comes from galaxy
cluster $M/L$ ratios. The most recent average
value $\Omega_{m} = 0.17 \pm 0.05$, has been determined by Bahcall and
Comerford \cite{Bahcall} by observing 21 galaxy clusters around $z
\sim 1$. The fact that nearby cluster data seem to prefer smaller
values for $\Omega_{m}$ than the one obtained from WMAP data at $z
\sim 1100$, can be regarded as a signal in favour of a decaying dark
matter model like the GCG.

Finally, one should expect a smaller ISW effect at early times, as the
gravitational potential is practically constant up to $z\simeq 2$ but,
at late times, there will be a larger ISW effect due to the energy
transfer from dark matter to dark energy resulting in a large
variation of the gravitational potential.


\section{Discussion and Conclusions}

In this work, we have considered a decomposition of the GCG in two
interacting components.  The first one can be regarded as dark matter
since it is pressureless. The second one has an equation of state,
$p_X = \omega_X \rho_X$. It has been shown that $\omega_X \le -1$.
Thus, once phantom-like behaviour is excluded our decomposition is
unique. This shows that in a unified model
 like GCG one does not need anyhting more than a simple cosmological 
constant as dark energy which is consistent with the recent study by Jassal 
et. al \cite{Hari} where it has been shown that a combination of WMAP 
data and observations of high redshift supernovae, is fairly consistent 
with a cosmological constant like dark energy.

It has been demonstrated that in this model the so-called dark
energy dominance is related with the time when matter fluctuations
become large ($\delta_{dm} > 1$).  Furthermore, we have shown that in
what concerns structure formation, the linear regime ($\delta_{dm}
\sim a$) is valid fairly close to the present, meaning that at the
time of structure formation begins, $z_c \simeq 10$, the dark energy
component was irrelevant and clustering occurred very much like in
the CDM model.  We have shown that the growth factor as well as the
bias parameter have a noticeable dependence on the $\alpha$
parameter. We have implemented a model where there is a violation of
the Equivalence Principle, as dark energy and baryons are not directly
coupled. This may turn out to be a distinct observational signature of
the present approach.

\begin{acknowledgments}

\noindent
M.C.B. and  O.B. acknowledge the partial support of FCT (Funda\c c\~ao para a 
Ci\^encia e a Tecnologia)
under the grant POCTI/FIS/36285/2000. The work of A.A. Sen is fully 
financed by FCT grant SFRH/BPD/12365/2003.

\end{acknowledgments}


\begin{thebibliography}{99}
\bibitem{Kamenshchik}  A. Kamenshchik, U. Moschella and V. Pasquier, \textit{%
Phys. Lett.} \textbf{511} (2001) 265.

\bibitem{Bento1}  M.C. Bento, O. Bertolami and A.A. Sen, \textit{Phys. Rev.} 
\textbf{D66} (2002) 043507.


\bibitem{Bento2} See e.g. M.C. Bento and  O. Bertolami,  \GRG {\bf  31} (1999)
1461; M.C. Bento, O. Bertolami and P.T. Silva, \PL {\bf B498} (2001) 62.



\bibitem{Ratra}  B. Ratra and P.J.E. Peebles, \textit{Phys. Rev.} \textbf{D37}
(1988) 3406; \textit{Ap. J. Lett.} \textbf{325} (1988) 117; C. Wetterich, 
\textit{Nucl. Phys.} \textbf{B302} (1988) 668; R.R. Caldwell, R. Dave and P.J.
Steinhardt, \textit{Phys. Rev. Lett.} \textbf{80} (1998) 1582; P.G.
Ferreira and M. Joyce, \textit{Phys. Rev.} \textbf{D58} (1998) 023503; I.
Zlatev, L. Wang and P.J. Steinhardt, \textit{Phys. Rev. Lett.} \textbf{82}
(1999) 986; P. Bin\'{e}truy, \textit{Phys. Rev.} \textbf{D60} (1999) 063502;
J.E. Kim, \textit{JHEP} 9905 (1999) 022; J.P. Uzan, \textit{Phys. Rev.} 
\textbf{D59} (1999) 123510; T. Chiba, \textit{Phys. Rev.} \textbf{D60}
(1999) 083508; L. Amendola, \textit{Phys. Rev.} \textbf{D60} (1999) 043501;
O. Bertolami and P.J. Martins, \textit{Phys. Rev.} \textbf{D61} (2000) 064007; 
\textit{Class. Quantum Gravity} \textbf{18} (2001) 593; A.A. Sen, S. Sen and S.
Sethi, \textit{Phys. Rev.} \textbf{D63} (2001) 107501; A.A. Sen and S. Sen, 
\textit{Mod. Phys. Lett.} \textbf{A16} (2001) 1303; A. Albrecht and C. Skordis, 
\textit{Phys. Rev. Lett.} 84 (2000) 2076; Y. Fujii, \textit{Phys. Rev.} 
\textbf{D61} (2000) 023504; M.C. Bento, O. Bertolami and N.C. Santos, \textit{
Phys. Rev.} \textbf{D65} (2002) 067301.


\bibitem{Bronstein}  M. Bronstein, \textit{Phys. Zeit. Sowejt Union} \textbf{%
3} (1938) 73; O. Bertolami, \textit{Il Nuovo Cimento} \textbf{93B} (1986)
36; \textit{Fortschr. Physik} \textbf{34} (1986) 829; 
M.Ozer and M.O Taha, \NP  \textbf{B287} (1987) 776.


\bibitem{Bento3}  M.C. Bento, O. Bertolami and A.A. Sen, \textit{Phys. Lett.}
\textbf{B575} (2003) 172; \textit{Phys. Rev.} \textbf{D67} (2003) 063003; 
\textit{Gen. Relativity and Gravitation} \textbf{35} (2003) 2063; D. Caturan
and F. Finelli, \textit{Phys. Rev.} \textbf{D68} (2003) 103501; Amendola, F.
Finelli, C. Burigana and D. Caturan, \textit{JCAP} \textbf{0307} (2003) 005.


\bibitem{Supern}  J.C. Fabris, S.B.V. Gon\c{c}alves and P.E. de Souza,
astro-ph/0207430; A. Dev, J.S. Alcaniz and D. Jain, \textit{Phys. Rev.} 
\textbf{D67} (2003) 023515; V. Gorini, A. Kamenshchik and U. Moschella, 
\textit{Phys. Rev.} \textbf{D67} (2003) 063509; M. Makler, S.Q. de Oliveira
and I. Waga, \textit{Phys. Lett.} \textbf{B555} (2003) 1; J.S. Alcaniz, D.
Jain and A. Dev, \textit{Phys. Rev.} \textbf{D67} (2003) 043514.


\bibitem{Silva}  P.T. Silva and O. Bertolami, \textit{Ap. J.} \textbf{599}
(2003) 829.


\bibitem{Bertolami}  O. Bertolami, A.A. Sen, S. Sen and P.T. Silva, \MNRAS to appear;
astro-ph/0402387.


\bibitem{Tonry} J.L. Tonry et al. \textit{Ap. J.} \textbf{594} (2003) 1.

\bibitem{Tritha}T. Roy Choudhury and T. Padmanabhan, astro-ph/0311622.

\bibitem{Bilic}  N. Bili\'{c}, G.B. Tupper and R.D. Viollier, \textit{Phys.
Lett.} \textbf{B535} (2002) 17; J.C. Fabris, S.B.V. Gon\c{c}alves and P.E. de Souza, 
\textit{Gen. Relativity and Gravitation} \textbf{34} (2002) 53.


\bibitem{Sandvik} H. Sandvik, M. Tegmark, M. Zaldarriaga and I. Waga, \PR \textbf{D69} (2004) 123524.


\bibitem{Cen} R. Cen, \textit{Ap. J.} \textbf{546} (2001) L77.

\bibitem{Paddy}T. Padmanabhan and T. Roy Choudhury \PR \textbf{D66} (2002) 081301.

\bibitem{Amendola1} L. Amendola, \PR \textbf{D62} (2000) 043511;
D. Tocchini-Valentini and L Amendola, \PR \textbf{D65} (2002) 063508.


\bibitem{Pavon} L.P. Chimento, A.S. Jakubi, D. Pav\'on and W. Zimdahl, \PR \textbf{D67} (2003) 083513.


\bibitem{Freeze} K. Freese, F.C. Adams, J.A. Frieman and E. Mottola, \NP \textbf{B287} (1987) 797.


\bibitem{Waga} R.C. Arcuri and I. Waga, \PR \textbf{D50} (1994) 2928.


\bibitem{Amendola2} L. Amendola and C. Quercellini, astro-ph/0403019.


\bibitem{Alam} U. Alam, V. Sahni, T.D. Saini and A.A. Starobinsky, 
astro-ph/0311364.


\bibitem{Bahcall} N. Bahcall and J.M. Comerford, \textit{Ap. J. Lett.} \textbf{565} (2002) L5.

\bibitem{Hari}H.K. Jassal, J.S. Bagla and T. Padmanabhan, astro-ph/0404378.


\end{thebibliography}
\end{document}